\documentstyle[aps]{revtex}

\begin{document}
\input epsf

\title{Double-logarithmic asymptotics of the magnetic form factor of electron
 and quark}

\author{B.I. Ermolaev\footnote{ Permanent address: A.F.Ioffe
Physico-Technical Institute,
194021 St.Petersburg,  Russia} \\
CERN, 1211 Geneva 23, Switzerland \\
and\\
S.I. Troyan\\
St.Petersburg Institute of Nuclear Physics, 188350
St.Petersburg-Gatchina, Russia}

\maketitle

\begin{abstract}
The asymptotical behaviour of the magnetic form factor for electron and
quark
is obtained in the double-logarithmic approximation for the Sudakov kinematics,
i.e. for the case when the value of the transfer momentum is much greater than
the mass of the particle
\end{abstract}

\section{Introduction}

The interaction of electron and quark with the
electromagnetic field is described in terms of two independent
form factors $f$ and $g$:

\begin{equation}
\Gamma_{\mu} = \bar{u}(p_2)[ \gamma_{\mu}f(q^2) - \frac{\sigma_{\mu
\nu}q_{\nu}}{2m}g(q^2)]u(p_1)
\label{1}
\end{equation}
where
$\sigma_{\mu \nu} = [\gamma_{\mu}\gamma_{\nu} - \gamma_{\nu}\gamma_{\mu}]/2$,
$q=p_2 - p_1$  is the momentum transferred to the electron or
the quark, $m$ is the electron or the quark mass and
both  the electric form factor $f$ and the magnetic form factor $g$ depend on
$q^2$.

In the Born approximation $f = 1$ and $g = 0$. One-loop
radiative corrections \cite{s} show that the form factor $f$ depends
on the infrared cut-off as well as on the ultraviolet one.
In contrast, the
form factor $g$ in the one-loop approximation is both ultraviolet-
and infrared-
stable. As the magnetic form factor $g$ contributes to  the
value of the anomalous magnetic momentum, it has been calculated
with great accuracy by direct
graph-by-graph calculations in the case when $q^2 = 0$. The most recent review
of such results is given in \cite{cm}.
Meanwhile, the electric form factor $f$ was calculated many years ago  in
the
 ``opposite'' kinematical
region  of very large transferred momenta :

\begin{equation}
                  -q^2 \gg m^2 ,
\label{2}
\end{equation}
in the leading logarithmic approximation (LLA), where the most important,
double logarithmic contributions to all orders in the QED coupling
$\alpha$ for the electron are taken into account.
The sum of such contributions, the double logarithmic (DL)
asymptotics for the electric form factor of the electron in the kinematical
region (\ref{2}) is

\begin{equation}
f = \exp\left[-\frac{\alpha}{4\pi}\ln^2(-q^2/m^2)\right]
\label{3}
\end{equation}

The famous expression (\ref{3}), obtained by V.V. Sudakov \cite{sud}
was actually the birth of
the approach that is so polular at present -- the
double logarithmic approximation (DLA) -- where only the leading contributions
$\sim (\alpha \ln^2(-q^2))^n$ are taken into account to all orders of the
perturbation series.  The generalization of the Sudakov form factor of
Eq. (\ref{3}) to quarks of QCD
obtained in \cite{sudqcd} amounts to replacing
$\alpha$ by $\alpha_s C_F$  in Eq. (\ref{3}),
where $C_F = (N^2-1)/2N = 4/3$ for the colour group $SU(3)$.

The exponential fall in Eq. (\ref{3}) as $-q^2$ increases corresponds to a
suppression of the non-radiative hard scattering of an electron by a virtual
photon.
The amplitude taking into account the
bremsstrahlung of $n$ ``soft'' photons
was shown in \cite{gorsh} to be the product of independent factors :

\begin{equation}
f_n =  B_1 B_2 ...B_n f(q^2) ,
\label{4}
\end{equation}
where the bremsstrahlung factors $B_i$  are given by (we drop the QED
coupling here)

\begin{equation}
B_ i =   \frac{p_2 l_i}{p_2k_i}  -\frac{p_1 l_i}{p_1 k_i}
\label{5}
\end{equation}
so that $l_i$ is the polarization vector and $k_i$ is the momentum of the $i$
-th
emitted photon $(i = 1, ...,n)$. As $f$ in Eq.~(\ref{5}) does not
depend on $k_i$, and each of
$B_i$ does not depend on $k_j$ with $j \neq i$,  Eq. (\ref{5})
leads to the Poisson energy spectrum for the bremsstrahlung photons
in the DLA.
The violation of the Poisson distribution in QCD for the emission of
the ``soft'' gluons in the DLA was obtained in \cite{fk} by calculating
the Feynman graphs up to order $\alpha_s^2$.
The generalization of the form factor $f$ in Eq. (\ref{4}) to
QCD was given in \cite{ef}, \cite{efl}.

In the present work we calculate the form factor $g$ for the electron and
the quark in the
kinematical region (\ref{2}) in the double logarithmic approximation.
Then, using results of \cite{efl} we obtain relations between
the radiative (inelastic) electric and magnetic
form factors of electron and quark.  Apart from the simplicity and beauty of
the expressions we obtain for the complete form of Eq. (1) in the
kinematical region of Eq. (\ref{2}), the reason for publication is that
those expressions
may be useful in the future for more precise descriptions of electron/quark
scattering. In particular, it can be applied to the analysis of electron
scattering off so-called magnetic walls, in the distant regions of our
Universe,
which together with  related phenomena are at present under discussion.
The paper is organized as follows: in Sect. 2 we calculate the magnetic
formfactor of the electron. In Sec. 2 we generalize that result to QCD.
Section 3 is devoted to concluding remarks.

\section{The magnetic form factor of the electron}

\begin{figure}
\begin{picture}(140,100)
\put(80,10){
\epsfxsize=3cm
\epsfysize=3cm
\epsfbox{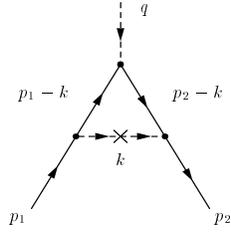}
}
\end{picture}
\caption{One-loop diagram for $\Gamma_{\mu}$.}
\label{oneloop}
\end{figure}

In the lowest, one-loop approximation, the magnetic form factor of the electron
was calculated long ago in \cite{s}. The only Feynman graph yielding
the main contribution to
$g$ in the kinematics (\ref{2}) is shown in Fig. \ref{oneloop}.
The result is \cite{s}

\begin{equation}
g^{(1)}(q^2) = -\frac{m^2}{q^2} \frac{\alpha}{\pi} \ln(-q^2/m^2) .
\label{a}
\end{equation}

In order to obtain the leading-log approximation to higher loop
contributions to the magnetic formfactor, we start by reproducing this
result in a way that will make it easy to generalize.

The Feynman diagram in Fig. \ref{oneloop} corresponds to the expression
\begin{equation}
\bar{u}_2\Gamma_{\mu}^{(1)}u_1 = e^2\int\frac{d^4k}{(2\pi)^4\imath}
\bar{u}_2\gamma_{\lambda}
\frac{m + \hat{p}_2 - \hat{k}}{[m^2 - (p_2 - k)^2 - \imath\epsilon]}
\gamma_{\mu}
\frac{m + \hat{p}_1 - \hat{k}}{[m^2 - (p_1 - k)^2 - \imath\epsilon]}
\gamma_{\tau} u_1 \frac{-g^{\lambda\tau}}{[-k^2 - \imath\epsilon]} .
\label{m1}
\end{equation}

The integral in Eq. (\ref{m1}) is both ultraviolet- and infrared-
divergent. To obtain a physically meaningful result, we must
introduce the infrared cut-off and subtract from Eq. (\ref{m1}) its
value at $q^2=0$.

As we want to obtain only the leading-log approximation (LLA) to the
resulting expression, we may use the approach of quasi-real photons
with the imposed cut-off $\mu$ on their transverse
momenta, \cite{l}. Choosing the cut-off parameter in the region

\begin{equation}
-q^2 \gg \mu^2 \gg m^2
\label{region}
\end{equation}
one can completely suppress the effects of electron mass, $m\to 0$, to
obtain the LLA result as a function of $\ln(-q^2/\mu^2)$. Finally,
we can shift $\mu^2$ down to $m^2$ to replace $\mu^2$ by $m^2$
in the obtained LLA expression.

Let us apply this procedure  to Eq. (\ref{m1}). In the approach of
quasi-real photons, the photon propagator is forced to be
on-the-mass-shell,

\begin{equation}
\frac{1}{-k^2 - \imath\epsilon} \to 2\pi\imath\delta^+(k^2) ,
\label{soh}
\end{equation}
and the quark virtualities are restricted by the largest virtuality in
the process:

\begin{equation}
2(p_2 k) \ll -q^2 ,\qquad 2(p_1 k) \ll -q^2 .
\label{restr}
\end{equation}

To perform the integrations we decompose the photon momentum in terms
of the initial and final quark momenta $p_1$ and $p_2$:

\begin{equation}
k = \alpha p_2 + \beta p_1 + k^{\perp} ,\qquad k^2 = \alpha\beta s +
(\alpha^2 + \beta^2)m^2 - k_{\perp}^2
\label{sudab}
\end{equation}
where $s=2(p_1p_2)=-q^2+2m^2\approx -q^2 \gg m^2$. Taking the integration
over $k_{\perp}$ with the help of the $\delta$-function  of Eq. (\ref{soh})
leads the remaining integral over $\alpha$, $\beta$ to have an integrand
symmetrical in $\alpha$ and $\beta$. As the
cut-off parameter $\mu$ is intentionally chosen to be in the region
(\ref{region}), we can neglect all terms $O(m^2/s)$ in the phase space

\begin{equation}
d^4k = d\alpha d\beta \frac{s}{2}d^2k_{\perp}
\label{phsp}
\end{equation}
and in the denominators of the quark propagators. In the case of the electric
form factor, we can also neglect the photon momentum $\hat{k}$ in
the spinor numerators of the propagators and immediately obtain the DLA
result:

\begin{equation}
f^{(1)}= - \frac{\alpha_e}{2\pi}\int^1\frac{d\alpha d\beta}{\alpha\beta}
\Theta\left(\alpha\beta - \frac{\mu^2}{s}\right)
\approx - \frac{\alpha_e}{4\pi}\ln^2\frac{-q^2}{\mu^2} .
\label{f1}
\end{equation}

To extract the magnetic part of the amplitude (\ref{m1}) we must not
neglect the mass $m$ and the photon momentum in the numerators and also
study the spinor structure of Eq. (\ref{m1}) in more detail.
Substituting Eq. (\ref{sudab}) in the numerators, and using the
commutation rules and the Dirac equation, we obtain

\begin{eqnarray}
& 2\bar{u}_2 [ s(1-\alpha)(1-\beta)\gamma_{\mu} +
m(p_1+p_2)_{\mu}(\alpha - \alpha^2 + \beta - \beta^2) +
\nonumber \\
& m q_{\mu}(\beta - \alpha + \beta^2 - \alpha^2) +
O(m^2, m k_{\perp}^2/s) ] u_1 .
\label{numer1}
\end{eqnarray}

Here the first term corresponds to the electric form factor. The second
term, which is proportional to

\begin{equation}
m \bar{u}_2 (p_1+p_2)_{\mu} u_1 =
m \bar{u}_2 \sigma_{\mu\nu}q^{\nu} u_1 + 2 m^2 \bar{u}_2\gamma_{\mu} u_1,
\label{sigm}
\end{equation}
determines the magnetic form factor. The third term, which might have
been violating the charge conservation, vanishes after integration over
$\alpha$, $\beta$ because of symmetry. As
$k_{\perp}^2\approx\alpha\beta s$~, we can neglect the terms
$O(k_{\perp}^2/s)$ as well as $O(\alpha\beta s)$ because they
suppress both logarithmic integrals over $\alpha$ and $\beta$.
Therefore for the one-loop contribution to the magnetic form factor we
obtain

\begin{equation}
g^{(1)} = - \frac{\alpha_e}{\pi} \frac{m^2}{q^2}
\int^1 \frac{d\alpha d\beta}{\alpha\beta}
[\alpha(1-\alpha) + \beta(1-\beta)] \Theta\left(\alpha\beta -
\frac{\mu^2}{s}\right)
\label{sigma}
\end{equation}
which reproduces the well-known result (\ref{a}). The large logarithmic
contribution to $g^{(1)}$ comes from two distinctly separate regions
of the phase space in the integral of Eq. (\ref{sigma}): one corresponds
to the ``hard''- photon radiated
in a collinear way to the momentum $p_1$ ($\beta\sim 1$, $\alpha \ll 1$) and
another -- collinearly to the momentum $p_2$ ($\alpha\sim 1$, $\beta \ll
1$).
The term ``hard''- photon radiation implies that, apart from a ``soft''-
photon radiation, one cannot ignore the recoil effect of the quark or
photon momentum in the spinor numerators of the quark
propagators.  When the photon is radiated along the momentum $p_1$, it
is the integral over the quark virtuality of the line $p_1$, $2(p_1
k)\approx\alpha s$, yields the log contribution, whereas the quark
virtuality of the line $p_2$, $2(p_2 k)\approx\beta s\sim -q^2$, is of
the largest scale.

For higher-loop diagrams in the approach of quasi-real photons, the LLA
contributions come from graphs with all photons emitted from the line
$p_1$ and absorbed on the line $p_2$. In the case of the electric form factor
$f$, the leading DL contributions come from ``soft''- photon emission when
each photon contributes to both logarithmic integrals over $\alpha$
and $\beta$. But for the case of the magnetic form factor $g$, one of these
photons turns out to be ``hard'' and collinear to $p_1$ or $p_2$.
We argue below that, as soon as other photons are to be ``soft'', each
yielding a DL contribution, their contributions to the magnetic
form factor turns out to be independent factors, similar to the case of
the electric form factor, i.e.

\begin{equation}
g^{(n)} = f^{(n-1)} g^{(1)} .
\label{gn}
\end{equation}

\begin{figure}
\begin{picture}(340,130)
\put(0,10){
\epsfbox{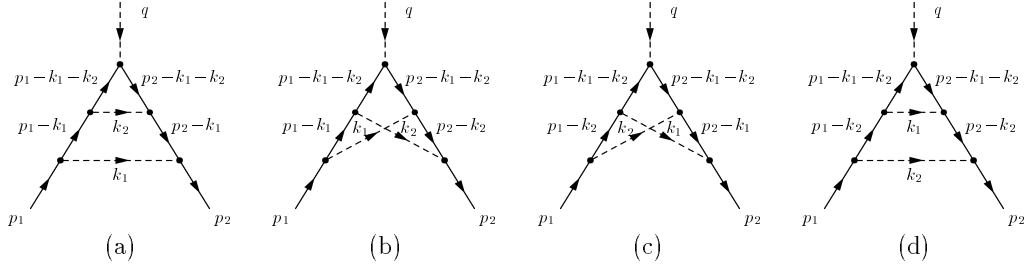}
}
\end{picture}
\caption{Two-loop diagrams for $\Gamma_{\mu}$ in DLA.}
\label{twoloop}
\end{figure}

Let us consider the two-loop diagrams with quasi-real photons in
Fig. \ref{twoloop}.
To prove this, we fix the momentum of the ``hard'' photon, $k_1$, to be
collinear to $p_1$ ($\beta_1\sim 1$, $\alpha_1 \ll 1$).  Then a
``soft'' photon $k_2$ can be emitted from the line $p_1$ either before
$k_1$ or later, as shown in Figs.   \ref{twoloop}a,b,c,d however, to provide a
DLA contribution, it must be absorbed on the line $p_2$ after $k_1$,
as shown in Figs. \ref{twoloop}b,d.
Indeed, the absorption part of the amplitude in the sum of two diagrams in
Figs. \ref{twoloop}d,c is

\begin{equation}
\bar{u}_2\left[
\gamma_{\lambda_2}
\frac{m\!+\!\hat{p}_2\!-\!\hat{k}_2}{m^2\!-\!(p_2\!-\!k_2)^2\!-\!\imath\epsilon}
\gamma_{\lambda_1} + \gamma_{\lambda_1}
\frac{m\!+\!\hat{p}_2\!-\!\hat{k}_1}{m^2\!-\!(p_2\!-\!k_1)^2\!-\!\imath\epsilon}
\gamma_{\lambda_2} \right]
\frac{m\!+\!\hat{p}_2\!-\!\hat{k}_1\!-\!\hat{k}_2}{m^2\!-\!(p_2\!-\!k_1\!-\!k_2)^2\!-\!\imath\epsilon}
\gamma_{\mu} \dots
\label{abso1}
\end{equation}

The above dots denote the remaining spinor part of the
amplitude of photon radiation from the quark line $p_1$.
Neglecting $\hat{k}_2$ in the spinor numerators and the $O(m^2/s)$ terms in
the denominators, and applying the commutation rules and the Dirac equation,
this expression in LLA in the region

\begin{equation}
\alpha_2,\/ \alpha_1 \ll 1 , \qquad\qquad \beta_2 \ll \beta_1 \sim 1
\label{rgn1}
\end{equation}
can be simplified to

\begin{equation}
\bar{u}_2 \left[ \left( \frac{2p_{2\lambda_2}}{\beta_2 s}
\gamma_{\lambda_1} + \gamma_{\lambda_1}
\frac{2(p_2-k_1)_{\lambda_2}}{\beta_1 s} \right)
\frac{m+\hat{p}_2-\hat{k}_1}{(\beta_2+\beta_1) s} +
\frac{\gamma_{\lambda_1}\gamma_{\lambda_2}}{(\beta_2+\beta_1) s}
\right] \gamma_{\mu} \dots
\label{abso2}
\end{equation}

Only the first term in Eq. (\ref{abso2}), which corresponds to
Fig. \ref{twoloop}d, provides
logarithmic integration over $\beta_2$ ; the other terms, which
correspond to Fig. \ref{twoloop}c with the ``soft'' $k_2$  photon
being absorbed before the ``hard'' $k_1$ photon, in can be
neglected in LLA.  Therefore expression (\ref{abso1}) in LLA acquires the
explicit factorized form:

\begin{equation}
\left(\frac{2p_{2\lambda_2}}{\beta_2 s} \right)
\bar{u}_2 \gamma_{\lambda_1}
\frac{m+\hat{p_2}-\hat{k}_1}{\beta_1 s} \gamma_{\mu} \dots
\label{abso3}
\end{equation}

The emission part of the amplitude in the sum of Figs. \ref{twoloop}d,b
providing the LLA contribution,

\begin{equation}
\dots \gamma_{\mu}
\frac{m\!+\!\hat{p}_1\!-\!\hat{k}_1\!-\!\hat{k}_2}{
m^2\!-\!(p_1\!-\!k_1\!-\!k_2)^2\!-\!\imath\epsilon}
\left[ \gamma_{\tau_1}
\frac{m\!+\!\hat{p}_1\!-\!\hat{k}_2}{
m^2\!-\!(p_1\!-\!k_2)^2\!-\!\imath\epsilon}
\gamma_{\tau_2} + \gamma_{\tau_2}
\frac{m\!+\!\hat{p}_1\!-\!\hat{k}_1}{
m^2\!-\!(p_1\!-\!k_1)^2\!-\!\imath\epsilon}
\gamma_{\tau_1} \right] u_1 ~,
\label{emit1}
\end{equation}
in the LLA region (\ref{rgn1}), turns into

\begin{equation}
\dots \gamma_{\mu}
\frac{m+\hat{p}_1-\hat{k}_1}{\alpha_1 s + (1-\beta_1)\alpha_2 s}
\left[ \gamma_{\tau_1} \frac{2p_{1\tau_2}}{\alpha_2 s} +
\frac{2(p_1-k_1)_{\tau_2}}{\alpha_1 s} \gamma_{\tau_1} \right] u_1 ~,
\label{emit2}
\end{equation}
where we again neglected $\hat{k}_2$ in the spinor numerators, all
$O(m^2/s)$ terms in the denominators and other terms leading beyond the
LLA. As the vector in square brackets is to be multiplied by the
polarization vector $p_{2\lambda_2}$ of expression (\ref{abso3}), we
can take the polarization vector $p_{1\tau_2}$ out of the brackets.
Although the denominators in expression (\ref{emit2}) provide log
integrations over $\alpha_1$ and $\alpha_2$ in two separate regions

\begin{equation}
\alpha_1 \ll (1-\beta_1)\alpha_2, \qquad (1-\beta_1)\alpha_2 \ll \alpha_1 ~,
\label{rgn2}
\end{equation}
the common numerator of the sum in square brackets exactly cancels the
largest denominator and turns expression (\ref{emit2}) into a
factorized amplitude of the independent emission of photons:

\begin{equation}
\dots \gamma_{\mu}
\frac{m+\hat{p}_1-\hat{k}_1}{\alpha_1 s} u_1 \left(
\frac{2p_{1\tau_2}}{\alpha_2 s} \right) .
\label{emit3}
\end{equation}

We would like to emphasize that deriving the factorized expressions
(\ref{abso3}), (\ref{emit3}), we did not neglect $m$ and any components
of $k_1$ in spinor numerators, which are essential for the magnetic
structure of the vertex $\Gamma_{\mu}$. Comparing expressions
(\ref{abso3})and (\ref{emit3}) with the one-loop amplitude (\ref{m1}), we
conclude that an additional ``soft'' photon loop does not spoil the spinor
structure and yields only an additional DL factor as in the case of
the electric form factor:

\begin{equation}
g^{(2)} = f^{(1)} g^{(1)} .
\label{g2}
\end{equation}

\begin{figure}
\begin{picture}(340,180)
\put(0,10){
\epsfbox{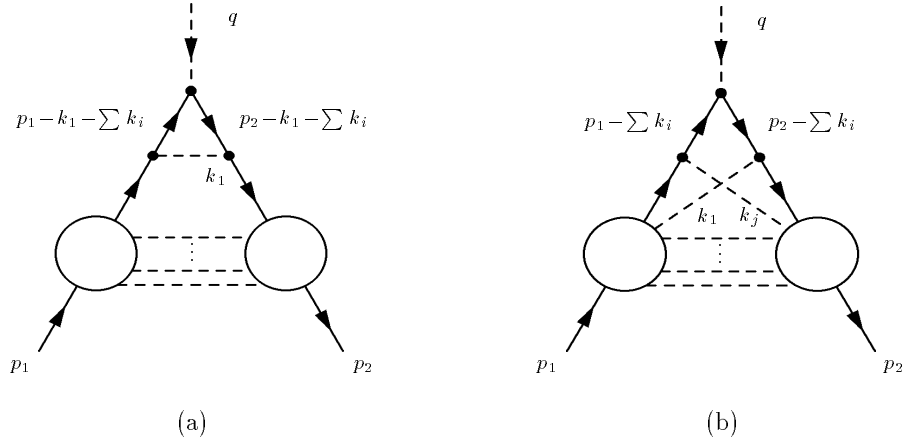}
}
\end{picture}
\caption{General graph for of $\Gamma_{\mu}$ in DLA.}
\label{manyloop}
\end{figure}

Now we are ready to submit arguments for Eq. (\ref{gn}) in a general
case. Consider the emission of $n$ quasi-real photons from the line $p_1$
and their absorption on the line $p_2$. Let us select one of them, e.g.
$k_1$, to be ``hard'' and collinear to $p_1$ ($\beta_1\sim 1$,
$\alpha_1 \ll 1$). This means that the photon $k_1$ must be absorbed
first, i.e. it is the one closest to the vertex $\gamma_{\mu}$ on  the
line $p_2$, as shown in Figs. \ref{manyloop}a,b, as its momentum $k_1$
introduces the largest possible virtuality for the quark line $p_2$:
$\beta_1 s\sim -q^2$.  Then the factorization property of absorption
amplitude of remaining ``soft'' photons on the final quark line $p_2$,
the blob on the line $p_2$ in Figs. \ref{manyloop}a,b,
is evident and well-known.  The same is true for the amplitude in
Fig. \ref{manyloop}a where all ``soft'' photons are emitted from
the incoming quark before its ``hard'' decay and where the recoil factor
$(1-\beta_1)$ does not influence the ``soft'' radiation in the blob
on the line $p_1$.  A more careful consideration is necessary for the
amplitude in Fig. \ref{manyloop}b, where ``soft'' photons are allowed to
radiate off the quark line $p_1$ as before the  ``hard'' one, $k_1$,
as later.

We prove the factorization property of the emission amplitude for a
general case with $n$ quasi-real photons by induction. Let us take for
granted that we have already proved the factorization property of the
emission amplitude in case of $(n-1)$ photons. Then monemtum $k_j$ of
the last photon emitted from the line $p_1$ in Fig. \ref{manyloop}b,
which is the one closest to the vertex $\gamma_{\mu}$ on the quark line
$p_1$, enters only the quark propagator between the vertices
$\gamma_{\lambda_j}$ and $\gamma_{\mu}$. One may consider, for a
moment,  $\gamma_{\lambda_j}$ to play the role of $\gamma_{\mu}$ and
use the factorization property of the remaining amplitude with $(n-1)$
photons:

\begin{equation}
\dots \gamma_{\mu}\gamma_{\tau_1}
\frac{m+\hat{p}_1-\hat{k}_1}{\alpha_1 s +
(1-\beta_1)(\alpha_2+\dots+\alpha_n) s} \gamma_{\tau_j}
\frac{m+\hat{p}_1-\hat{k}_1}{\alpha_1 s} u_1
\prod^{n}_{\stackrel{\scriptstyle i=2}{i\neq j}}
\left( \frac{2p_{1\tau_i}}{\alpha_i s} \right) .
\label{emitn}
\end{equation}

Again we have neglected here all ``soft'' $\hat{k}_i$ in the numerators and
the $O(m^2/s)$ terms in denominators. Let us apply to this expression
the commutation rules and Dirac equation, and take into account the fact that
the polarization vector of the $k_j$ photon must be multiplied to
the polarization vector $p_2^{\tau_j}$ of its absorption in the
blob on the line $p_2$ in Fig. \ref{manyloop}b. Summing over all
``soft'' photons we get

\begin{equation}
\sum^{n}_{j=2}
\!\frac{(1\!-\!\beta_1)2p_{1\tau_j}}{[\alpha_1 \!+\!
(\!1\!-\!\beta_1)(\!\alpha_2\!+\!\dots\!+\!\alpha_n)]s} \frac{1}{\alpha_1
s} \!\left(\!\prod^{n}_{\stackrel{\scriptstyle i=2}{i\neq j}}
\frac{2p_{1\tau_i}}{\alpha_i s} \!\right) \!=\!
\prod^{n}_{i=2} \!\left(\!
\frac{2p_{1\tau_i}}{\alpha_i s} \!\right)
\frac{[(\!1\!-\!\beta_1)(\!\alpha_2\!+\!\dots\!+\!\alpha_n)]s}{[\!\alpha_1 \!+\!
(\!1\!-\!\beta_1)(\!\alpha_2\!+\!\dots\!+\!\alpha_n\!)]s}\frac{1}{\alpha_1 s} .
\label{esumb}
\end{equation}

Adding this expression to the evident resulting factor for the graph
in Fig. \ref{manyloop}a,

\begin{equation}
\prod^{n}_{i=2} \left(
\frac{2p_{1\tau_i}}{\alpha_i s} \right)
\frac{1}{[\alpha_1 +
(1-\beta_1)(\alpha_2+\dots+\alpha_n)]s} .
\label{esuma}
\end{equation}

we finally prove that the ``soft'' photons
contribution in DLA is just a scalar factor to the spinor
structure of the one-loop graph, which is the same for both the
electric and magnetic form factors.

Summing over the number of photons $n$ then leads to the final
generalization of the DLA relation Eq. (\ref{gn}) of magnetic and
electric form factors:

\begin{equation}
g = f g^{(1)} ,\qquad f = \sum^{\infty}_{n=0} f^{(n)} = \exp(f^{(1)}) .
\label{gf}
\end{equation}

Taking into account the explicit expressions for $f^{(1)}$ and
$g^{(1)}$, Eqs. (\ref{f1}) and (\ref{a}), this relation can be read as

\begin{equation}
g = -2 \frac{\partial}{\partial \rho} f
\label{gdf}
\end{equation}
with $\rho=s/\mu^2\approx -q^2/m^2$ (as $\mu^2$ is shifted down to
$m^2$). For Eq. (\ref{1}) we therefore obtain in DLA
the following formula:

\begin{equation}
\bar{u}_2\Gamma_{\mu}u_1 = \bar{u}_2\left[ \gamma_{\mu} +
\frac{\sigma_{\mu \nu} q_{\nu}}{m} \frac{\partial}{\partial \rho}\right]
u_1 \exp\left[-\frac{\alpha}{4\pi} \ln^2\rho~\right]
\label{gamma}
\end{equation}
or, in a different form,

\begin{equation}
\bar{u}_2\Gamma_{\mu}u_1 = \bar{u}_2\left[ \gamma_{\mu} - \frac12
m\sigma_{\mu\nu} \frac{\partial}{\partial q_{\nu}}\right]
e^{-\frac{\alpha}{4\pi} \ln^2(-q^2/m^2)} u_1
\label{intrig}
\end{equation}

\section{The magnetic form factor of the quark}

The reason for the simplicity of  relation (\ref{gdf}) between $f$ and $g$,
which we obtained in the previous
section, is that they differ in only one respect:
compared to $f$, $g$ misses
only one logarithmic contribution coming  from any of
the photon propagators attached to the uppermost vertices
in Fig. \ref{manyloop}.
Otherwise, they are identical. So, $g$  can be regarded
as the result
of a convolution of the single-logarithmic first-loop contribution
with the infinite number of double-logarithmic contributions
from the other loops. Such a picture is helpful to get a generalization
of (\ref{gamma}) to QCD.
The main technical difference between calculating the form factor $g$
for electron and for quark is that, in QCD, the three-gluon vertices
also are essential in the DLA, increasing considerably the number
of involved Feynman graphs in each order in $\alpha_s$.
However,  they contribute to both $f$ and $g$, and the result for
$f_q$, electric form factor of the quark, is \cite{sudqcd}

\begin{equation}
f_q = \exp[-\frac{\alpha_s C_F}{4\pi} \ln^2\rho~ ] .
\label{fq}
\end{equation}
where $C_F =(N^2 - 1)/ 2N = 4/3$ .

In other words, the whole effect of replacement of the electromagnetic
gauge group $U(1)$ by $SU(3)$ results in the replacement of
$\alpha$ in Eq. (\ref{gdf})
by $\alpha_s C_F ~(C_F =(N^2 - 1)/ 2N = 4/3)$
as the DL contributions of diagrams with three-gluon vertices turn to
cancel each other in the total sum of graphs for the amplitude (\ref{1}).
For the case of $g_q$ form factor this observation was made in two-loop
calculation in \cite{br}. In order to prove the exponentiation of DL
radiative corrections for the electric and magnetic form factors, $f_q$ and
$g_q$, one might follow the approach developed in \cite{efl}, which is
based on generalization of the Gribov bremsstrahlung theorem \cite{g} to
QCD, for the case of the electric form factor.
Repeating that proof here would lead us far beyond the scopes of the
present paper. Thus we  restrict ourselves just with the statement
that with all orders in
$\alpha_s$ taken into account, we arrive at the following expression for
the vertex $\Gamma^{(q)}_{\mu}$ of the quark in the external
electromagnetic field in the
kinematical region (\ref{2}) of large transfer momenta:

\begin{equation}
\Gamma^{(q)}_{\mu} = \left[ \gamma_{\mu} +
\frac{\sigma_{\mu \nu} q_{\nu}}{m} \frac{\partial}{\partial \rho}\right]
\exp\left[-\frac{\alpha_s C_F}{4\pi} \ln^2\rho~\right] ~.
\label{quark}
\end{equation}

\section{Conclusion}

We have calculated the magnetic form factor of
electron and quark in the asymptotical regime where the momentum $q$
transferred to the electron or the quark is much greater than their mass.
Eqs.~(\ref{gamma}) and (\ref{quark}) that
we have obtained predict  an exponential fall for the magnetic form factors
when $q^2$ increases. This corresponds
to the suppression of non-radiative scattering, without
photon bremsstrahlung,  at high energies.
When the photon bremsstrahlung is taken into account in the DLA,
in the expressions for such radiative (inelastic) form factors
Eqs.~(\ref{gamma}) and (\ref{quark})
are multiplied  by the bremsstrahlung
factors $B_i$ defined in Eq. (\ref{5}).
All the form factors in Eqs. (\ref{gamma}) and (\ref{quark})
do not depend on the features of the emitted photons.
 It means that taking into account the magnetic formfactor
does not violate the Poisson energy spectrum for the photon
bremsstrahlung in the
double logarithmic approximation. On the contrary, the inelastic electric
form factor that accounts for the gluon bremsstrahlung depends, in the DLA,
both on the
emitted gluon momenta and on the structure of each of the gluon cascades
\cite{ef,efl}. This makes possible to obtain the simple expressions for
the inelastic formfactor in QCD only separately for every kinematical
region fixed by a certain ordering of the emission energies and
angles \cite{ef}. Still, in the DLA, even with that complication, the
relation (\ref{gdf}) holds also for the inelastic electric and magnetic
formfactors.

\section{acknowledgements}

We are grateful to G. Altarelli, M. Mangano, N. Nikolaev and D. Soper for
useful discussions.

\newpage

\begin{thebibliography}{99}

\bibitem{s} J.Schwinger. Phys.~Rev. 73 (1948) 416.
\bibitem{cm} A.Czarnecki and W.J.Marciano. BNL-HET-98/43; hep-ph/9810512.
\bibitem{sud} V.V.Sudakov. ZhETP 3 (1956) 65.
\bibitem{sudqcd} J.J.Carazzone, E.C.Poggio and  H.R.Quinn.
Phys.~Rev. ~D11 (1975) 2286;
J.M.Cornwall and G.Tiktopoulos. Phys.~Rev.~D13 (1976) 3370.
\bibitem{gorsh} V.G.Gorshkov. ZhETP 56 (1969) 598.
\bibitem{fk} V.S.Fadin and E.A.Kuraev. Sov.~J.~Nucl.~Phys. 27 (1983) 587.
\bibitem{ef} B.I.Ermolaev and V.S.Fadin.JETP Lett. 33 (1981) 269;
   V.S.Fadin. Sov.J.Nucl.Phys. 37 (1983) 2145.
\bibitem{efl} B.I.Ermolaev, V.S.Fadin and L.N.Lipatov.
Sov.~J.~Nucl.~Phys. 45 (1987) 508.
\bibitem{l} L.N.Lipatov. Phys.~Lett.~B116 (1982) 411.
\bibitem{br} R.Barbieri and E.Remiddi. Nuovo Cimento 11A (1972) 824.
\bibitem{g} V.N.Gribov. Sov. J. Nucl. Phys. 5 (1967) 280.

\end{thebibliography}
\end{document}